\documentclass[aps,pre,showpacs,preprint]{revtex4}

\usepackage{amsmath}
\usepackage{amssymb}
\usepackage[utf8]{inputenc}
\usepackage{graphicx}
\usepackage{dcolumn}
\usepackage[mathscr]{euscript}
\usepackage{bm}
 
\DeclareMathOperator\erfi{erfi}

\begin{document}

\title{Kinetic model for a confined quasi-two-dimensional gas of inelastic hard spheres}
\author{J. Javier Brey, P. Maynar, and M. I. Garc\'{\i}a de Soria}
\affiliation{F\'{\i}sica Te\'{o}rica, Universidad de Sevilla,
Apartado de Correos 1065, E-41080, Sevilla, Spain \\ E-mail: brey@us.es}

\date{\today }

\begin{abstract}

The local balance equations for the density, momentum, and energy of a dilute gas of elastic or inelastic hard spheres,  strongly confined between two parallel hard plates are obtained.  The starting point is a Boltzmann-like  kinetic equation, recently derived for this system. As a consequence of the confinement, the pressure tensor and the heat flux contain, in addition to the terms associated to the motion of the particles,   collisional transfer contributions, similar to those that appear beyond the dilute limit. The complexity of these terms, and of the kinetic equation itself, compromise the potential of the equation to describe the rich phenomenology observed in this kind of systems. For this reason, a simpler model equation based on the Boltzmann equation is proposed. The model is formulated to keep the main properties of the underlying equation,  and it is expected to provide relevant information in more general states than the original equation. As an illustration, the solution describing a macroscopic state with uniform temperature, but a density gradient  perpendicular to the plates is considered. This is the equilibrium state for an elastic system, and the inhomogeneous cooling state for the case of inelastic hard spheres. The results are in good agreement  with previous results  obtained directly from the Boltzmann equation. 

\end{abstract}

\maketitle

\section{Introduction}
\label{s1}
Fluids in extreme confinement between two parallel plates, have many peculiar properties as compared with the behavior of  bulk systems far away from the boundaries.
This includes both equilibrium \cite{TGyR92,DyH95,RSLyT97,FLyS12} and non-equilibrium situations  \cite{KyD89,LBOHFyS10} of molecular gases, and also  granular gases, i.e. systems composed of macroscopic particles whose interactions are inelastic \cite{Metal05,MyS16}.  For instance, in this latter case, it has been experimentally observed  that when energy is continuously injected into the system to compensate the energy dissipation in collisions, the system reaches a spatially homogeneous steady state for a wide range of values of the density. Nevertheless, for certain densities, which depend on the values of the magnitudes characterizing the energy injection, the final state reached by the system exhibits two coexisting phases, one of them having a smaller number of particles density and a larger kinetic energy density than the other \cite{PEyU02,RCByH11,CMyS12,CMyS15}. Although several effective models have been proposed to describe this phenomenology with more or less success \cite{BRyS13,CLyH00,BBGyM16,RSyG18,MGyB19a},  a satisfactory explanation, based on a well established   description of the dynamics of the system is still lacking. One of the main reasons for this is that the form of the macroscopic hydrodynamic equations describing the evolution of the system is unknown. For strongly confined systems, the effect of the physical boundaries occurs not only through the boundary conditions to be imposed to the macroscopic evolution equations, but it is expected to modify also the structure of the transport equations themselves. The changes will be deeper  if the dynamics observed by projecting on a plane parallel to the plates is considered, and   the interest focusses in formulating macroscopic transport equations for that quasi-two-dimensional dynamics. 

Kinetic theory provides an intermediate level of description of a system of particles in which the details and basis for macroscopic balance equations have their origins. To isolate the more important distinguished features of the extreme confinement, the simplest case of inelastic smooth hard spheres at low density is considered here. The bulk macroscopic hydrodynamic equations for a system composed of these particles have been extensively investigated, their structure has been established and explicit expressions for the coefficients appearing in them has been obtained. The  theoretical predictions following  from them   agree quite well with molecular dynamics simulations results \cite{BDKyS98,ByC01,Go03}. The starting point for this research was the Boltzmann kinetic equation modified to account for inelastic two-particle collisions. From it, exact balance equations were derived for the mass, momentum, and energy. Then, to close those equations, a modified Chapman-Enskog method was employed to obtain a ``normal solution''  of the kinetic equation as an expansion in spatial gradients of the macroscopic fields. The analysis is something more complex that in the case of elastic particles, due to the inelasticity of collisions that implies that there is no equilibrium state.  Instead, a particular solution describing a homogeneous state in which the temperature decreases monotonically in time (the so-called homogeneous cooling state), was used as the  reference state to carry out the expansion. 

In order to implement a similar program for strongly confined systems, the first issue is to formulate a kinetic equation describing the dynamic of the system. Recently, a Boltzmann-like kinetic equation has been derived for a system of hard spheres confined between two infinite parallel plates at rest, separated a distance smaller that two particles diameters.  The equation can be applied, to both elastic \cite{BMyG16,BGyM17} and inelastic particles \cite{MGyB19,BGyM19a}. The only difference is in the collision rule used to determine the change of the velocities  when two particles collide. For elastic systems, the kinetic equation has been proven to obey an H theorem, implying the tendency of any solution to a final equilibrium state. This state coincides with the one predicted by equilibrium statistical mechanics: the velocity distribution is Maxwellian with a uniform temperature and the density profile is inhomogeneous along the direction perpendicular to the plates. In the case of inelastic hard spheres, if no external energy is continuously injected into the system, trivially there is no equilibrium state. Instead there is again a particular state showing a uniform temperature that decreases monotonically in time, similarly to what happens in the homogeneous cooling state. On the other hand, as in the equilibrium state, the density is not uniform along the direction perpendicular to the plates. For this reasons, this state has been referred to as the inhomogeneous cooling state \cite{BGyM19a}

In this work, balance equations for the density, momentum, and energy are derived from the Boltzmann equation for the confined system. This is done in the more general case of inelastic hard spheres, but the results for the elastic case are easily obtained by taking the appropriate limit. As a consequence of the extreme confinement, the pressure tensor and the heat flux have, in addition to the kinetic contributions associated to the motion of the particles, collisional transfer contributions that, in the present case, are due to the restriction on the allowed collisions imposed by the boundaries. An analysis of the structure of the collisional transfer contributions leads to a result that generalizes the ``contact theorem'' of equilibrium statistical mechanics \cite{HyMc06} to arbitrary non-equilibrium states and also to inelastic particles for extreme confinement. Actually, this  is a particular case of a completely general property following from the dynamics itself of  confined hard spheres \cite{MGyB18}.

The complexity of the collision term makes quite difficult to apply the kinetic equation to realistic problems, beyond situations for which a linearization around the reference state  (equilibrium or the ICS) can be done. For this reason, it is useful to formulate a kinetic model that permits analytical studies in fully nonlinear problems. A model kinetic equation is obtained by replacing the collision term with a much simpler form that preserves the most important properties. These include normalization and the local conservation laws for the number of particles, momentum, and energy. By preserving the equations, it is meant that, not only the structure of the equations is the same, but also that the expressions giving the pressure tensor and the heat flux, and in the inelastic  case the energy sink term, are the same functionals of the distribution function as in the original description. The model proposed here is inspired in the well-known Bhatnagar-Gross-Kook  (BGK) model for the usual Boltzmann equation \cite{Ce75}. The BGK model has been extensively used and studied. Its limitations are well established, but also the means to correct them are known. As an illustration, the solution of the model kinetic equations describing the ICS  is investigated and the results compared with the description provided by the original Boltzmann equation. The agreement can be considered as quite satisfactory.

The remaining of the paper is organized as follows. In the next section, le Boltzmann equation for the confined system of hard spheres is reviewed.  The local balance laws  are derived from it by taking velocity moments. The collisional transfer contributions to the pressure tensor and the heat flux, due to the extreme confinement, are identified.  The model kinetic equation is formulated in Sec.\, \ref{s3}, where its relation with the original Boltzmann equation is stressed. As an initial relevant test of the model, it is shown to have a solution describing the same macroscopic ICS as the original equation, namely uniform decreasing temperature and density gradient along the direction perpendicular to the plates. The comparisson of the cooling rates and the density profiles shows a good qualitative and quantitative agreement. The final section of the paper contains a short summary and some analysis of the possible applications of the model. Details of the calculations are provided in the Appendices.

\section{description of the system and balance equations}
\label{s2}
\begin{figure}
\includegraphics[scale=0.8,angle=0]{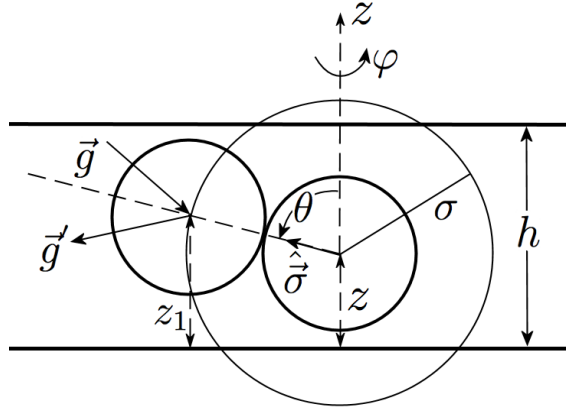}
\caption{Coordinates used for the description of the collision of two hard spheres of diameters $\sigma$,  confined in a quasi-two-dimensional system by means of two hard parallel plates separated a distance $h$,  $\sigma< h< 2 \sigma$.}
\label{fig1}
\end{figure}

The Boltzmann equation, describing the time evolution of the one-particle distribution function, $f ({\bm r},{\bm v},t)$ for a dilute gas of inelastic hard spheres of mass $m$ and diameter $\sigma$, confined between two large hard parallel plates separated a distance $h$ smaller than  twice the diameter of a particle, $\sigma<h < 2 \sigma$, has the form \cite{BMyG16,BGyM17,MGyB19}
\begin{equation}
\label{2.1}
\frac{\partial f}{\partial t}+{\bm v} \cdot \frac{\partial f}{\partial {\bm r}}= J[{\bm r},{\bm v}|f],
\end{equation}
with thw binary collision term given by
\begin{eqnarray}
\label{2.2}
J[{\bm r},{\bm v}|f] & = & \sigma  \int d{\bm v}_{1} \int_{0}^{2\pi} d \varphi \int_{\sigma/2}^{h-\sigma/2} dz_{1}\,   | {\bm g} \cdot \widehat {\bm \sigma} | \nonumber \\
&& \times | \left[ \Theta ({\bm g} \cdot \widehat {\bm \sigma})  \alpha^{-2} b_{\bm \sigma}^{-1} - \Theta (-{\bm g} \cdot \widehat{\bm \sigma})\right]f(x,y,z_{1},{\bm v}_{1} ,t) f({\bm r},{\bm v},t).
\end{eqnarray}
Here ${\bm r}$ is the position vector of components  $\{x,y,z \}$, $ {\bm r}_{1z} \equiv \{x,y,z_{1} \}$, ${\bm g} \equiv {\bm v}_{1}-{\bm v}$, $\Theta$ is the Heaviside step function, and $b_{\bm \sigma}^{-1}$ is an operator changing all the velocities ${\bm v}$ and ${\bm v}_{1} $ to its right into their pre-collisional values for a collision defined by the unit vector $\widehat{\bm \sigma}$,
\begin{equation} 
\label{2.3}
{\bm v}^{*} \equiv b_{\bm \sigma}^{-1} {\bm v} = {\bm v} + \frac{1 + \alpha}{2 \alpha} \left( {\bm g} \cdot \widehat{\bm \sigma} \right) \widehat{\bm \sigma},
\end{equation}
\begin{equation} 
\label{2.4}
{\bm v}^{*}_{1} \equiv b_{\bm \sigma}^{-1} {\bm v}_{1} = {\bm v} -  \frac{1 + \alpha}{2 \alpha} \left( {\bm g} \cdot \widehat{\bm \sigma} \right) \widehat{\bm \sigma}.
\end{equation}
In the above expressions, and also in Eq.\ (\ref{2.2}), $\alpha$ is the coefficient of normal restitution, defining the inelasticity of collisions and defined in the interval $0<\alpha \leq 1$.
The value $\alpha=1$ corresponds to the limit of elastic collisions. The unit vector $\widehat{\bm \sigma}$ is given by
\begin{equation}
\label{2.5}
\widehat{\bm \sigma} \equiv  \left\{ \sin \theta \sin \varphi, \sin \theta \cos  \varphi, \cos \theta \right\}
\end{equation}
with $\varphi$ being an azimuth angle and 
\begin{equation}
\label{2.6}
\cos \theta = \frac{z_{1}-z}{\sigma}, \quad \sin \theta\geq 0 .
\end{equation}
A scheme of the coordinates used to describe the collision of two particles is given in Fig. \ref{fig1}.

The kinetic equation (\ref{2.1}) holds for $\sigma/2 < z< h-\sigma/2$ and it has to be solved with the appropriate boundary conditions \cite{BGyM17}. These conditions guarantee that if the initial one-particle distribution function vanishes outside the system, this property is kept in time by the solution of the kinetic equation. For an arbitrary function $\phi ({\bm v})$, the collision term has the useful property
\begin{eqnarray}
\label{2.7}
\int d{\bm v}\,  \phi ({\bm v}) J[{\bm r},{\bm v}| f]   &= & \sigma \int d{\bm v} \int d{\bm v}_{1} \int_{\sigma/2}^{h-\sigma/2} dz_{1}   \int_{0}^{2\pi} d \phi\, |  {\bm g} \cdot \widehat{\bm \sigma}| \Theta (-{\bm g} \cdot \widehat{\bm \sigma})  \nonumber \\
&& \times  f({\bm r}_{1z},{\bm v},t) f({\bm r},{\bm v},t) \left( b_{\bm \sigma} -1\right) \phi({\bm v}),
\end{eqnarray}
where $b_{\bm \sigma}$ is the inverse operator of $b_{\bm \sigma}^{-1}$, i.e. it changes all the velocities ${\bm v}$ and ${\bm v}_{1}$ to its right into the post-collisional values given by
\begin{equation}
\label{2.8}
 {\bm v}^{\prime} \equiv b_{\bm \sigma} {\bm v}= {\bm v} + \frac{1+\alpha}{2} \left( {\bm g} \cdot \widehat{\bm \sigma} \right) \widehat{\bm \sigma},
\end{equation}
\begin{equation}
\label{2.9}
 {\bm v}^{\prime}_{1} \equiv b_{\bm \sigma} {\bm v}_{1} = {\bm v}_{1}- \frac{1+\alpha}{2} \left( {\bm g} \cdot \widehat{\bm \sigma} \right) \widehat{\bm \sigma}.
\end{equation}
Macroscopic density, $n({\bm r},t)$, velocity, ${\bm u}({\bm r},t)$,  and granular temperature, $T({\bm r},t)$  fields are defined in the usual way,
\begin{equation}
\label{2.10}
n({\bm r},t) \equiv \int d{\bm v}, f({\bm r}, {\bm v},t),
\end{equation}
\begin{equation}
\label{2.11}
n({\bm r},t) {\bm u}({\bm r},t) \equiv \int d{\bm v}\, {\bm v} f({\bm r},{\bm v},t),
\end{equation}
\begin{equation}
\label{2.12}
\frac{3}{2} n({\bm r},t) T({\bm r},t) \equiv  \frac{m}{2} \int d{\bm v}\, V^{2}({\bm r},t) f({\bm r},{\bm v},t).
\end{equation}

Here, ${\bm V} ({\bm r},t) \equiv {\bm v}-{\bm u},({\bm r},t)$ is the peculiar velocity of the particle  relative to the local flow velocity. Notice that, as it is usual in the literature of granular gases, the Boltzmann constant does not appear in the definition of the granular temperature. This is done to emphasize the absence of any thermodynamic meaning in such definition. From Eq. (\ref{2.1})), balance equations can be derived for the macroscopic fields by taking velocity moments. They have the form
\begin{equation}
\label{2.13}
\frac{\partial n({\bm r},t)}{\partial t} + \frac{\partial}{\partial {\bm r}} \cdot \left[ n({\bm r},t) {\bm u} ({\bm r},t )\right] =0,
\end{equation}
\begin{equation}
\label{2.14}
m n({\bm r},t) \frac{\partial {\bm u} ({\bm r},t)}{\partial t} + m n ({\bm r},t) {\bm u}({\bm r},t) \cdot  \frac{\partial {\bm u}({\bm r},t)}{\partial {\bm r}} + \frac{\partial}{\partial {\bm r}} \cdot {\sf P}({\bm r},t)=0,
\end{equation}
\begin{eqnarray}
\label{2.15}
\frac{3}{2} n({\bm r},t) \frac{\partial T({\bm r},t)}{\partial t} & + & \frac{3}{2} n({\bm r},t)) {\bm u}({\bm r},t) \cdot \frac{\partial T({\bm r},t)}{\partial {\bm r}} +{\sf P}({\bm r},t) : \frac{\partial {\bm u}({\bm r},t)}{\partial {\bm r}} \nonumber \\
&+& \frac{\partial}{\partial {\bm r}}\, \cdot {\bm J_{q}}({\bm r},t) = \left(1-\alpha^{2} \right) \omega ({\bm r},t).
\end{eqnarray}
The pressure tensor ${\sf P}({\bm r},t)$, and the heat flux ${\bm J}_{q}({\bm r},t)$ have both ``kinetic'' and ``collisional transfer'' contributions,
\begin{equation}
\label{2.16}
{\sf P}({\bm r},t) = {\sf P}^{(k)}({\bm r},t) + {\sf P}^{(c)}({\bm r},t),
\end{equation}
\begin{equation}
\label{2.17}
{\bm J}_{q}({\bm r},t)= {\bm J}_{q}^{(k)}({\bm r},t) + {\bm J}_{q}^{(c)}({\bm r},t).
\end{equation}
The kinetic contributions are given by
\begin{equation}
\label{2.18}
{\sf P}^{(k)}({\bm r},t)= m\int d{\bm v}\, {\bm V}({\bm r},t) {\bm V} ({\bm r},t) f({\bm r},{\bm v},t),
\end{equation}
\begin{equation}
\label{2.19}
{\bm J}_{q}^{(k)} ({\bm r},t)) = \frac{m}{2} \int d{\bm v}\, V^{2}({\bm r},t) {\bm V}({\bm r},t) f({\bm r},{\bm v},t).
\end{equation}
The collisional transfer parts of the fluxes verify
\begin{equation}
\label{2.20}
\frac{\partial}{\partial z} {\sf P}_{zi}^{(c)}({\bm r},t) = - m  \int d{\bm v}\,  v_{i} J[{\bm r},{\bm v}|f],
\end{equation}
\begin{equation}
\label{2.21}
\frac{\partial}{\partial z} J_{q,z}^{(c)}({\bm r},t)= - \frac{m}{2} \int d{\bm v} v^{2} J[{\bm r},{\bm v} |f] - \frac{\partial}{\partial z}  \left[ {\bm u}({\bm r},t) \cdot {\sf P}^{(c)}({\bm r},t) \right] -(1-\alpha^{2}) \omega ({\bm r},t)
\end{equation}
Using  these relations, the following explicit expressions can be identified
\begin{eqnarray}
\label{2.22}
{\sf P}_{ij}^{(c)} ({\bm r},t) &=& \delta_{iz} \frac{1+\alpha}{4}\, \sigma^{2} m \int d{\bm v} \int d {\bm v}_{1} \int_{\sigma/2}^{h-\sigma/2} d z_{1} \int_{0}^{2\pi} d \varphi \int_{0}^{1} d\lambda\, \lambda^{-1} |{\bm g} \cdot \widehat{\bm \sigma}_{\lambda}|^{2}\widehat{\sigma}_{\lambda z} \widehat{\sigma}_{\lambda j}  \nonumber \\
& & \times \Theta \left( -{\bm g} \cdot \widehat{\bm \sigma}_{\lambda} \right) f({\bm r}_{1z},{\bm v}_{1},t)) f({\bm r}_{\lambda z},{\bm v},t),
\end{eqnarray}
\begin{eqnarray}
\label{2.23}
J_{qi}^{(c)}({\bm r},t)& = & \delta_{iz} \frac{1+\alpha}{4}\, \sigma^{2} m \int d{\bm v} \int d {\bm v}_{1} \int_{\sigma/2}^{h-\sigma/2} d z_{1} \int_{0}^{2\pi} d \varphi \int_{0}^{1} d\lambda\, \lambda^{-1} |{\bm g} \cdot \widehat{\bm \sigma}_{\lambda}|^{2}\widehat{\sigma}_{\lambda z} \cal{\bm G}\cdot \widehat{\bm \sigma}_{\lambda} \nonumber \\
& & \times \Theta \left( -{\bm g} \cdot \widehat{\bm \sigma}_{\lambda} \right) f({\bm r}_{1z},{\bm v}_{1},t)) f({\bm r}_{\lambda z},{\bm v},t).
\end{eqnarray}
Here,
\begin{equation}
\label{2.24}
\mathcal{\bm G}({\bm r},t) \equiv  \frac{{\bm V}({\bm r},t)+ {\bm V}_{1} ({\bm r},t)}{2} = \frac{{\bm v}+{\bm v}_{1}}{2}\, - {\bm u}({\bm r},t),
\end{equation}
i.e. it is the center of mass velocity of the colliding particles relative to the macroscopic flow. Moreover
\begin{equation}
\label{2.25}
{\bm r}_{\lambda z} \equiv \left\{ x,y,z_{\lambda} \right\},
\end{equation}
\begin{equation}
\label{2.26}
z_{\lambda} \equiv  \frac{z-(1-\lambda)z_{1}}{\lambda},
\end{equation}
and
\begin{equation}
\label{2.27}
\widehat{\bm \sigma}_{\lambda} \equiv \left\{ \sin \theta_{\lambda} \sin \varphi, \sin \theta_{\lambda} \cos \varphi, \cos \theta_{\lambda} \right\},
\end{equation}
where
\begin{equation}
\label{2.28}
\cos \theta_{\lambda} = \frac{z_{1}-z}{\lambda \sigma}\, ,  \quad \sin \theta_{\lambda} = + \sqrt{1- \cos^{2} \theta_{\lambda}}\, .
\end{equation}

Finally, the source term on the right hand side of the energy balance equation, Eq.\, (\ref{2.15}), describes the kinetic energy dissipation in collisions due to inelasticity, and it can not be written as the divergence of a flux. Its expression is
\begin{equation}
\label{2.29}
\omega ({\bm r},t )= \frac{m \sigma}{8} \int d{\bm v} \int d {\bm v}_{1} \int_{\sigma/2}^{h-\sigma/2} d z_{1} \int_{0}^{2\pi} d \varphi\,  |{\bm g}  \cdot \widehat{\bm \sigma} |^{3} \Theta  \left( - {\bm g}  \cdot \widehat{\bm \sigma} \right) f({\bm r}_{1z},{\bm v}_{1},t) f({\bm r},{\bm v},t).
\end{equation}

Some details of the derivation of the balance laws and the expressions of the fluxes and the source term are given in Appendix \ref{ap1}.  The collisional transfer contributions to the pressure tensor and the heat flux are due to  the delocalization of the centres of the colliding pair of particles and, more specifically, to the different values of the two $z$ coordinates. This is  a consequence of the confinement of the system that restricts the possible values of the collision vector. The explicit form of this restriction for a particle, depends on the vertical coordinate of the particle  and, for this reason, the vertical coordinates of the pair of colliding particle are relevant in describing the dynamics of the system even at low densities.  When considering the   expressions of the collisional transfer contributions to the pressure tensor and to the heat flow, Eqs. (\ref{2.22}) and (\ref{2.23}), it must be kept in mind that that, as already indicated, for physical initial conditions, the one-particle velocity distribution, $f({\bm r},{\bm v},t)$ vanishes for positions outside the system, i.e. for both $z<\sigma/2$ and $z>h-\sigma/2$. As a consequence,  the effective range of the $\lambda$-integration is restricted to values such that
\begin{equation}
\label{2.30}
\frac{\sigma}{2} < \frac{z-(1-\lambda)z_{1}}{\lambda} <h-\frac{\sigma}{2}\, .
\end{equation}
This eliminates the apparent divergence of the integrands for $\lambda=0$, and also guarantees that $-1 \leq \cos \theta_{\lambda} \leq 1$.  The previous comment is closely related with an important general property of a system of hard spheres or disks confined by means of hard walls. Consider $z=\sigma/2$, so that
\begin{equation}
\label{2.31}
z_{\lambda}= \frac{\sigma/2 -(1-\lambda)z_{1}}{\lambda}\, .
\end{equation}
For $z_{1}=\sigma/2$, it is $z_{\lambda} (z=\sigma/2,z_{1}=\sigma/2)= \sigma/2$. Besides 
\begin{equation}
\label{2.32}
\left( \frac{\partial z_{\lambda}}{\partial z_{1}} \right)_{z} = - \frac{1-\lambda}{\lambda}\,  ,
\end{equation}
that is negative for $0<\lambda<1$. It follows that, as $z_{1}$ increases from $\sigma/2$, $z_{\lambda}$ decreases from the same vale. Let is stress that this holds for the particular value $z=\sigma/2$. Consequently, $z_{\lambda}$ always takes values outside the system for $z=\sigma/2$, and $f({\bm r}_{\lambda z},{\bm v},t)=0$, implying that
\begin{equation}
\label{2.33}
{\sf P}_{ij}^{(c)}({\bm r},t)=0
\end{equation}
and
{\begin{equation}
\label{2.34}
{\bm J}_{q}^{(c)} ({\bm r},t)=0
\end{equation}
for $z=\sigma/2$. A similar argument shows that the collisional transfer contributions to the pressure tensor and the heat flux also vanish at the other plate, i.e. for $z=h-\sigma/2$. The property that the collisional part of the pressure vanishes at a hard wall for a system of hard spheres or disks at equilibrium is known asas the ``contact theorem'' \cite{HyMc06}. Recently, it has been extended to arbitrary non-equilibrium states and shown that, actually,  all the components of the collisional transfer contribution to the pressure tensor vanish \cite{MGyB18}. Here, it has been shown that the same property applies in the case of a strongly confined gas of inelastic hard spheres, and that the collisional transfer contribution to the heat flux also vanishes at the hard boundary. 

\section{Kinetic model}
\label{s3}
The Boltzmann equation (\ref{2.1}) is technically more involved than the usual Boltzmann equation for bulk systems, due to the space dependence of the collision term. For this reason, it is appealing to consider kinetic equations obtained as approximated representations of the original Boltzmann equation, that allow a more detailed analytical study. The basic idea of a kinetic model is to look for the maximum simplicity while retaining the  most relevant physical and mathematical properties of the original equation. In the present case, these include the existence of the equilibrium state in the elastic limit and the exact form of the conservation laws. This will accomplished by considering the collision term $J$ as expanded in a complete set of velocity polynomials with a scalar product weighted with the local equilibrium distribution function. The contribution from the subspace leading to the balance equations for the macroscopic fields  is kept exactly, while the   rest is approximated by a single relaxation term. Similar models have been successfully formulated for the revised Enskog theory of a non-confined system of hard spheres, both for elastic \cite{DSyB96,SMDyB98} and dissipative  \cite{BDyS97} collisions.
To derive the kinetic model, it is useful to introduce a Hilbert space of real functions of the velocity ${\bm v}$ by means of the scalar product
\begin{equation}
\label{3.1}
\langle g|h\rangle \equiv \int d{\bm v}\, \psi_{l}({\bm v} ) g({\bm v}) h({\bm v}),
\end{equation}
where $\psi _{l}({\bm v})$ is the Maxwellian defined with the actual values of the local velocity and granular temperature at position ${\bm r}$ and time $t$,
\begin{equation}
\label{3.2}
\psi _{l}({\bm v}) \equiv \left[ \frac{m}{2 \pi T({\bm r},t)} \right]^{3/2} \exp - \frac{mV^{2}({\bm r},t)}{2 T ({\bm r},t)}\, .
\end{equation}
This function is related with the usual local equilibrium distribution function $f_{l}({\bm r},{\bm v},t) $ by
\begin{equation}
\label{3.3}
f_{l}({\bm r}, {\bm v},t ) = n({\bm r},t) \psi_{l}({\bm v}).
\end{equation}
Notice that both $\psi_{l}$ and $f_{l}$ depend on position and time through their functional dependence on the macroscopic fields, although it is not explicitly indicated for the former. In this sense, the Hilbert space is defined for given values of ${\bm r}$ and  $t$, namely those corresponding to the position and time at which the distribution function will be evaluated.
Consider next the set of functions
\begin{equation}
\label{3.4}
\left\{ \phi_{\beta} \right\} \equiv \left\{ 1, \left( \frac{m}{T} \right)^{1/2} {\bm V}, \left( \frac{2}{3} \right)^{1/2} \left( \frac{mV^{2}}{2T}-\frac{3}{2} \right) \right\}.
\end{equation}
These functions are orthogonal withe the scalar product definition in Eq.\ (\ref{3.1}),
\begin{equation}
\label{3.5}
\langle \phi_{\beta} | \phi_{\beta^ \prime} \rangle = \delta_{\beta, \beta \prime}.
\end{equation}
In the following the two notations
\begin{equation}
\label{3.6}
\left\{ \phi_{\beta} \right\} \equiv \left\{ \phi_{1},{\bm\phi}_{2}, \phi_{5} \right\}
\end{equation}
and
\begin{equation}
\label{3.7}
\left\{ \phi_{\beta} \right\} \equiv \left\{ \phi_{1},\phi_{2}, \phi_{3},\phi_{4},\phi_{5} \right\}
\end{equation}
will be employed indistinctly. A projection operator $\mathcal{P}$ over the subspace spanned by the functions $\phi_{\beta}$ is defined by
\begin{equation}
\label{3.8}
\mathcal{P} h({\bm v}) \equiv \psi_{l}({\bm v}) \sum_{\beta} \phi_{\beta}({\bm v}) \langle \phi_{\beta} | \psi_{l}^{-1} h \rangle = \psi_{l}({\bm v}) \sum_{\beta} \phi_{\beta} ({\bm v})  \int d {\bm v}^{\prime} \phi_{\beta}({\bm v}^{\prime}) h({\bm v}^{\prime}).
\end{equation}
By means of this operator, the collision term $J$ in the kinetic equation (\ref{2.1}) is decomposed into the two parts
\begin{equation}
\label{3.9}    
J[{\bm r},{\bm v}|f]  =\mathcal{P} J[{\bm r},{\bm v}|f] + (1-\mathcal{P} )J[{\bm r},{\bm v}|f] .
\end{equation}
It is
\begin{equation}
\label{3.10}
\int d{\bm v}\, \phi_{\beta}({\bm v}) \mathcal{P} h({\bm v}) = \int d{\bm v} \phi_{\beta}({\bm v}) h({\bm v}),
\end{equation}
for arbitrary $h({\bm v})$ and, consequently,
\begin{equation}
\label{3.11}
\int d{\bm v}\, \phi_{\beta}({\bm v})  \left( 1- \mathcal{P} \right) J[{\bm r},{\bm v}|f]=0.
\end{equation}
Then, the second term on the right hand side of Eq.\ (\ref{3.9}) does not contributes to the form of the balance equations. Here it will be approximated by a single exponential relaxation term,
\begin{equation}
\label{3.12}
(1-\mathcal{P} ) J[{\bm r},{\bm v}|f] \rightarrow  -\nu \left[ f({\bm r},{\bm v},t) - f_{l}({\bm r}, {\bm v}, t) \right],
\end{equation}
where $\nu$ is a velocity independent characteristic frequency that can be a functional of the density and granular temperature fields. It can be chosen to optimize the agreement between the kinetic model and the original Boltzmann equation with regards to some property of interest. From the definition of the local equilibrium distribution it follows that
\begin{equation}
\label{3.13}
\int d{\bm v}\, \phi_{\beta}({\bm v}) f({\bm r},{\bm v}.t) = \int d{\bm v}\, \phi_{\beta}({\bm v}) f_{l}({\bm r},{\bm v},t),
\end{equation} 
$\beta=1,\ldots,5$, that implies
\begin{equation}
\label{3.14}
\mathcal{P} \left[-\nu (f-f_{l}) \right] =0,
\end{equation}
showing the consistency of the approximation made in Eq.\ (\ref{3.12}).  Then, the model kinetic  equation reads
\begin{equation}
\label{3.15}
\frac{\partial}{\partial t}\, f({\bm r},{\bm v},t) +{\bm v} \cdot \frac{\partial}{\partial {\bm r}}\, f({\bm r},{\bm v},t) = -\nu \left[ f({\bm r},{\bm v},t)-f_{\l}({\bm r},{\bm v},t) \right] + \mathcal{P} J[{\bm r},{\bm v}|f].
\end{equation}
By construction, this model equation leads to the same balance equations as the original Boltzmann equation (\ref{2.1}), since the part of the collision term that has been approximated does not contribute to those equations. The term $\mathcal{P} J$ can be written in a more explicit way in terms of the collisional transfer contributions to the fluxes, ${\sf P}^{(c)}$ and ${\bm J}_{q}^{(c)}$, and the source term, $\omega$. By using Eqs. (\ref{2.20}) and (\ref{2.21}), the resulting kinetic equation is
\begin{equation}
\label{3.16}
\frac{\partial f}{\partial t} + {\bm v} \cdot \frac{\partial f}{\partial {\bm r}} =  -\nu \left( f-f_{l} \right)- \frac{f_{l}}{nT}  \left[ V_{i}   \frac{\partial}{\partial z} {\sf P}_{zi}^{(c)} + \left( \frac{mV^{2}}{3T}-1 \right) 
\left( \frac{\partial}{\partial z}\, J_{q,z}^{(c)} - (1-\alpha^{2}) \omega + {\sf P}_{zi}^{(c)} \frac{\partial u_{i}}{\partial z} \right) \right],
\end{equation}
where repeated indices are implicitly summed over. This equation is a highly non-lineal integro-differential equation. The local equilibrium distribution, the collisional transfer contributions of the pressure tensor and the heat flux, and the source energy term appearing in the equation, are given by the same functional of $f$ as in the original Boltzmann equation, i.e. by Eqs.  (\ref{2.22}), (\ref{2.23}), and (\ref{2.29}), respectively. The main and essential mathematical advantage of Eq. (\ref{3.16}) as compared with the original  Boltzmann equation (\ref{2.1}), is that the velocity dependence of the collision term on the right hand side is much simpler now. It is given by a Gaussian times polynomials of second degree. A relevant example of application of the model is given in the next section.
 
 \section{The inhomogeneous cooling state}
 \label{s4}
 A special idealized state of the confined system we are considering here is that of inhomogeneous cooling state (ICS), for which all the time dependence of the macroscopic dynamics occurs through the time dependence of the granular temperature, while there is a spatial dependence through the density, that is a function of the distance to the confining walls \cite{BGyM19a}. The ICS occurs in freely evolving systems, i.e. without any kind on external energy injection. It  is an extension to confined quasi-two-dimensional granular gases of the homogeneous cooling state exhibited by bulk systems \cite{Ha83}.  
 
 The model kinetic equation (\ref{3.16}) has a solution, $f_{ICS}$, whose macroscopic description corresponds to the ICS. It is given by the Maxwellian 
 \begin{equation}
 \label{4.1}
 f_{ICS}(z,{\bm v},t) = n(z) \psi_{ICS}({\bm v},t)
 \end{equation}
 with
 \begin{equation}
 \label{4.2}
 \psi_{ICS}({\bm v},t)= \left[ \frac{m}{2\pi T(t)} \right]^{3/2} \exp - \frac{mv^{2}}{2T(t)}\, .
 \end{equation}
 The density profile must satisfy the equation
 \begin{equation}
 \label{4.3}
 \frac{\partial}{\partial z} \ln n(z) = \pi (1+\alpha) \int_{\sigma/2}^{h-\sigma/2} dz_{1}\, (z-z_{1}) n(z_{1})
 \end{equation}
 and the time evolution of the granular temperature is given by
 \begin{equation}
 \label{4.4}
 \frac{\partial}{\partial t}\, T(t) = -\frac{2}{ 3 n(z)}\ (1-\alpha^{2}) \omega.
 \end{equation}
 The proof of the above results is given in Appendix \ref{ap2}. Using Eq. (\ref{4.1}) into Eq. (\ref{2.29}) it is easily obtained 
 \begin{equation}
 \label{4.5}
 \omega(z,t) =  \left( \frac{\pi}{m} \right)^{1/2} \frac{N \sigma}{A}\, n(z) T(t)^{3/2}\, ,
 \end{equation}
 so that Eq.\ (\ref{4.4}) can be expressed in the usual for
 \begin{equation}
 \label{4.6}
 \frac{\partial}{\partial t}\, T(t) =- \zeta(t) T(t),
 \end{equation}
 with the cooling rate given by
 \begin{equation}
 \label{4.7}
 \zeta(t) = \frac{2}{3}\, \left( 1 -\alpha^{2} \right) \frac{N \sigma}{A} \left[ \frac{\pi T(t)}{m} \right]^{1/2}\, .
 \end{equation}
 In these expression s$A$ is the area of each of the two plates confining the system, so that $N/A$ can be seen as an effective two-dimensional number of particles density. As required by consistency with the existence of the ICS, the cooling rate does not depend on $z$. 
 
 Let us consider the equation for the density profile, Eq. (\ref{4.3}).  Taking into account that for symmetry considerations it must be $n(z)=n(h-z)$, the equation is seen to be equivalent to
 \begin{equation}
 \label{4.8}
 \frac{\partial}{\partial z}\, \ln n(z) = \frac{\pi N}{A} (1+\alpha) \left( z-\frac{h}{2} \right),
 \end{equation}
 whose solution is
 \begin{equation}
 \label{4.9}
 n(z) = \frac{N}{Ab} \exp \left[ a \left(z-\frac{h}{2} \right)^{2}\right]
 \end{equation}
 where
 \begin{equation}
 \label{4.10}
 a \equiv \frac{\pi (1+\alpha) N}{2A}\, ,
 \end{equation}
 \begin{equation}
 \label{4.11}
 b \equiv \left( \frac{\pi}{a}\right)^{1/2}  \erfi \sqrt{a} \left( \frac{h-\sigma}{2} \right).
 \end{equation}
 Here $\erfi(x)$ is the imaginary error function defined as
 \begin{equation}
\label{4.12}
\erfi(y) \equiv \pi^{-1/2} \int_{-y}^{y} dy^{\prime}\, e^{y^{\prime2}}.
\end{equation}
 In the elastic limit $\alpha=1$, the above density profile coincides with the exact equilibrium result reported in ref. \cite{BMyG16}. For $\alpha<1$, the comparison can be made with an approximate result derived from the Boltzmann equation (\ref{2.1}) also for the ICS, and valid in the limit $(h-\sigma)/\sigma \ll 1$. The only difference is that the latter contains a factor due to the anisotropy of the velocity distribution function, an effect that is neglected in the model kinetic equation. Let us mention that the inhomogeneity of the density field in the direction perpendicular to the confining plates, and the quantitative accuracy of the theoretical predictions based on the Boltzmann equation have been verified both for  elastic \cite{BMyG16} and inelastic  \cite{BGyM19a} hard spheres by means of molecular dynamics simulations.
 
 Since, by construction,  the model leads to the same balance equation as the original Boltzmann equation, Eqs.\ (\ref{2.13})-(\ref{2.15}) can be directly applied to analyze some properties of the ICS, as predicted by the model. From the momentum conservation equation, it follows that the component of the pressure tensor in the ICS must verify
 \begin{equation}
 \label{4.13}
 \frac{\partial {\sf P}_{zx}^{(c)}}{\partial z} = \frac{\partial {\sf P}_{zy}^{(c)}}{\partial z} =0,
 \end{equation}
 \begin{equation}
 \label{4.14}
 \frac{\partial {\sf P}_{zz}}{\partial z} = \frac{\partial}{\partial z}\ \left( {\sf P}_{zz}^{(k)} + {\sf P}_{zz}^{(c)} \right) =0.
 \end{equation}
 Since the model alkso conserves the expresion of the pressure tensor, the general result derived in Sec. \ref{s2} that the collisional treansfer contributions to the pressure tensor vanish at the hard walls can be employed, and write
 \begin{equation}
 \label{4.15}
 {\sf P}_{zz}(t)= {\sf P}_{zz}^{(k)} (z= \sigma/2,t) = n(z=\sigma/2) T(t),
 \end{equation}
 where Eq.\ (\ref{ap2.1}) has been used. This expression gives the normal  force per unit of area exerted by the gas on the plates in the ICS. In the elastic limit, the only change to be made is tu substitute the time dependent temperature by its equilibrium value. Then, the expression agrees with the result obtained previously for the equilibrium state of an elastic gas of hard spheres \cite{SyL97,BMyG16}. The simplicity of this result is consistent with the behaviour obtained in the limit of extreme confinement also for elastic spheres \cite{FLyS12}. It is worth to emphasize that, as derived here, Eq. (\ref{4.15}) reflects a property valid not only at the hard boundaries of the system but also in the bulk.

 The distribution function of the ICS following from the Boltzmann equation  (\ref{2.1})  differs qualitatively from a Maxwellian, specially for $\alpha$ significantly smaller than unity,  while we have seen that with the kinetic model a Maxwellian is obtained for all values of $\alpha$. In spite of this difference, the kinetic model may still be useful to study  a variety of physical situations. To put this expectation on a firmer basis, the velocity distribution function of the ICS has been computed using molecular dynamics simulation techniques. The results show that the velocity distribution is anisotropic, i.e. the marginal distributions associated to the velocity parallel and perpendicular to the plates differ, and so do the associated partial granular temperatures. On the other hand, both distributions can be approximated very accurately by a Maxwellian for thermal velocities, say $|v|  \lesssim (T/m)^{1/2}$, even for relatively small values of the coefficient of normal restitution and for widths of the system close to $2 \sigma$.

 \section{Summary and Conclusions}
 \label{s4}
 
 One of the aims of this paper is to derive the balance equations for the local density, momentum, and energy of a system of (elastic or inelastic) hard spheres  confined between two infinite parallel plates, separated a distance smaller than two particle diameters. The dynamics of the system is assumed to be accurately described by a Boltzmann-like equation derived recently.  As a consequence of the confinement, there are collisional transfer contributions to the pressure tensor and the heat flux that, in principle, are characteristic of fluid beyond the dilute limit.  Quite interestingly, the expressions of the collisional transfer contributions imply that all them vanish at the hard boundaries confining the system. This result applies independently of the state of the system and also of the nature, elastic or inelastic, of collisions. Therefore, it constitutes a wide generalization of the ``wall theorem'' formulated in equilibrium statistical mechanics. Another aim of the present work is to formulate a model kinetic equation closely related to the Boltzmann-like kinetic equation mentioned above.  In particular, the model has, by construction, the same balance equations as the original equation. As a test of the model, it has been applied to the simpler macroscopic states corresponding to a molecular, elastic gas and to a granular, inelastic gas, respectively.  A good agreement with previous results obtained using  equilibrium methods and the Boltzmann equation itself has been found.
 
 In principle, the model formulated here could be used to derive hydrodynamic transport equations by using the Chapman- Enskog procedure.  Actually, this has been done for other BGK-like models formulated for the Enskog equation of dense gases of hard spheres \cite{DSyB96,BDyS97}. Nevertheless, this may be not a good strategy to describe the observed macroscopic behavior exhibited by this system and mentioned in the Introduction section.  The observed phenomenology corresponds to a two-dimensional image of the system, when looked from above or from below. The effective phase separation occurs   in the plane parallel to the plates and not in a volume region of the system. Moreover, and more important, the geometry of the system makes it difficult to suppose that there is hydrodynamic behaviour in the direction perpendicular to the plates.. It seems more plausible  that the dynamical behavior observed by projecting over a horizontal plane can be accurately described at a macroscopic level by hydrodynamic-like equations. In this context, a logical way to proceed is to derive projected balance equations from  the three-dimensional ones presented here, formulating afterwards a model kinetic equation using arguments similar to those used in this paper.

 \acknowledgments

This research was supported by the Ministerio de Econom\'{\i}a, Industria  y Competitividad  (Spain) through Grant No. FIS2017-87117-P (partially financed by FEDER funds).

 \appendix

\section{The macroscopic balance equations}
\label{ap1}
Muiplication of Eq. (\ref{2.1}) by $m{\bm v}$ and integration over ${\bm v}$ yields
\begin{equation}
\label{ap1.1}
mn({\bm r},t) \frac{\partial {\bm u}({\bm r},t)}{\partial t} + m n({\bm r},t) {\bm u}({\bm r},t) \cdot \frac{\partial {\bm u}({\bm r}.t)}{\partial {\bm r}} + \frac{\partial}{\partial {\bm r}}\, \cdot {\sf P}^{(k)}({\bm r},t) = \int d{\bm v}\, m {\bm v} J[{\bm r},{\bm v}|f],
\end{equation}
 with the kinetic parte of the pressure tensor, ${\sf P}^{(k)}$, given by Eq.\ (\ref{2.18}). By using the property (\ref{2.7}) and the collision rule, Eq. (\ref{2.8}), it is found
 \begin{eqnarray}
 \label{ap1.2}
 \int d{\bm v}\, m {\bm v} J[{\bm r},{\bm v}|f] & = & - \frac{1+\alpha}{2}\ m \sigma \int d{\bm v} \int d {\bm v}_{1} \int_{\sigma/2}^{h-\sigma/2} dz_{1} \int_{0}^{2\pi} d \varphi\,  |{\bm g} \cdot \widehat{\bm \sigma} |^{2} \widehat{\bm \sigma} \Theta (-{\bm g} \cdot \widehat{\bm \sigma}) \nonumber \\
 && \times f({\bm r}_{1z},{\bm v}_{1} ,t) f({\bm r},{\bm v},t).
\end{eqnarray}
This expression is trivially equivalent to
\begin{eqnarray}
 \label{ap1.3}
 \int d{\bm v}\, m {\bm v} J[{\bm r},{\bm v}|f] & = & - \frac{1+\alpha}{2}\ m \sigma \int d{\bm v} \int d {\bm v}_{1} \int_{\sigma/2}^{h-\sigma/2} dz_{1} \int_{\sigma /2}^{h-\sigma/2} dz_{2}  \int_{0}^{2\pi} d \varphi\,  \delta (z-z_{2}) |{\bm g} \cdot \widehat{\bm \sigma}_{2} |^{2}  \nonumber \\
 && \times  \widehat{\bm \sigma}_{2} \Theta (-{\bm g} \cdot \widehat{\bm \sigma}_{2})  f({\bm r}_{1z},{\bm v}_{1} ,t) f({\bm r}_{2z},{\bm v},t),
\end{eqnarray}
 with
 \begin{equation}
 \label{ap1.4}
 \widehat{\bm \sigma}_{2} \equiv \left\{  \sin \theta_{2} \sin \varphi, \sin \theta_{2} \cos \varphi, \cos \theta_{2} \right\},
 \end{equation}
 \begin{equation}
 \label{ap1.5}
 \cos \theta_{2} = \frac{z_{1}-z_{2}}{\sigma}, \quad \sin \theta_{2} >0,
 \end{equation}
 and
 \begin{equation}
 \label{ap1.6} 
 {\bm r}_{2z} \equiv \left\{ x,y,z_{2} \right\}\, .
 \end{equation}
 It is worth to remark that for the equivalence of Eqs. (\ref{ap1.2}) and (\ref{ap1.3}) it is crucial that $\sigma/2 < z< h/2$. Interchange of $z_{1}$ and $z_{2}$,  of ${\bm v}_{1}$ and ${\bm v}_{2}$, and change of $\varphi$ into $\varphi+\pi$ on the right-hand-side of Eq. (\ref{ap1.3}) leads to
 \begin{eqnarray}
 \label{ap1.7}
 \int d{\bm v}\, m {\bm v} J[{\bm r},{\bm v}|f] & =  &\frac{1+\alpha}{2}\ m \sigma \int d{\bm v} \int d {\bm v}_{1} \int_{\sigma/2}^{h-\sigma/2} dz_{1} \int_{\sigma /2}^{h-\sigma/2} dz_{2}  \int_{0}^{2\pi} d \varphi\,  \delta (z-z_{1}) |{\bm g} \cdot \widehat{\bm \sigma}_{2} |^{2}  \nonumber \\
 && \times  \widehat{\bm \sigma}_{2} \Theta (-{\bm g} \cdot \widehat{\bm \sigma}_{2})  f({\bm r}_{1z},{\bm v}_{1} ,t) f({\bm r}_{2z},{\bm v},t),
\end{eqnarray}
where we have taken into account that under the indicated changes $\widehat{\bm \sigma}_{2}$ changes signe. Taking one half of each of the two equivalent expressions given by Eqs. (\ref{ap1.3}) and (\ref{ap1.7}), one gets
\begin{eqnarray}
\label{ap1.8}
\int d{\bm v}\, m {\bm v} J[{\bm r},{\bm v}|f] & =  & -\frac{1+\alpha}{4}\ m \sigma \int d{\bm v} \int d {\bm v}_{1} \int_{\sigma/2}^{h-\sigma/2} dz_{1} \int_{\sigma /2}^{h-\sigma/2} dz_{2}  \int_{0}^{2\pi} d \varphi\, \left[  \delta (z-z_{2})  \right.\nonumber \\ 
&& \left.- \delta (z-z_{1}) \right]|{\bm g} \cdot \widehat{\bm \sigma}_{2} |^{2}  
   \widehat{\bm \sigma}_{2} \Theta (-{\bm g} \cdot \widehat{\bm \sigma}_{2})  f({\bm r}_{1z},{\bm v}_{1} ,t) f({\bm r}_{2z},{\bm v},t).
\end{eqnarray}
Next, the formal relation
\begin{equation}
\label{ap1.9}
\delta (z-z_{2})-\delta (z-z_{1}) = -(z_{2}-z_{1}) \frac{\partial}{\partial z} \int_{0}^{1} d\lambda\, \delta \left[ z-\lambda (z_{2}-z_{1}) -z_{1} \right]
\end{equation}
is used, so that Eq.\ (\ref{ap1.8}) is seen to be equivalent to
\begin{eqnarray}
\label{ap1.10}
\int d{\bm v}\, m {\bm v} J[{\bm r},{\bm v}|f] & =  & -\frac{1+\alpha}{4}\ m \sigma^{2}  \frac{\partial}{\partial z} \int d{\bm v} \int d {\bm v}_{1}  \int_{\sigma/2}^{h-\sigma/2} dz_{1} \int_{\sigma /2}^{h-\sigma/2} dz_{2}  \int_{0}^{2\pi} d \varphi \int_{0}^{1} d\lambda\,\,  \nonumber \\ 
&& \times \delta \left[ z-\lambda (z_{2}-z_{1})-z_{1} \right]  {\bm g} \cdot \widehat{\bm \sigma}_{\lambda} |^{2}  
   \widehat{\bm \sigma}_{2} \Theta (-{\bm g} \cdot \widehat{\bm \sigma}_{\lambda})  \nonumber \\
   && \times  f({\bm r}_{1z},{\bm v}_{1} ,t) f({\bm r}_{\lambda z},{\bm v},t). 
 \end{eqnarray}
 In the above expression, ${\bm r}_{\lambda z}$ and  $\widehat{\bm \sigma}_{\lambda}$ are given by Eqs. (\ref{2.25}) and (\ref{2.27}), respectively. The presence of the delta function in the integrand guarantees that $-1 < \cos \theta_{\lambda} <1$, as it must be. Consider
 \begin{equation}
 \label{ap1.11}
 \int_{\sigma/2}^{h-\sigma/2} dz_{2}\,  \delta \left[ z-\lambda \left/z_{2}-z_{1} \right)-z_{1} \right] = \int_{\sigma/2}^{h-\sigma/2} \delta \left[ \lambda \left( z_{2}- z_{\lambda} \right) \right].
 \end{equation}
 This integral is equal to $\lambda^{-1}$ for $\sigma/2 <z_{\lambda} <h-\sigma/2$ and vanishes otherwise. Taking into account that $f({\bm r},{\bm v},t)$ also  vanishes for $z$ outside this interval, one easily gets  Eq.\ (\ref{2.20}) from Eq. (\ref{ap1.10}), with  the collisional transfer contribution to the pressure tensor given by Eq.\, (\ref{2.22}). Finally, use of Eq.\, (\ref{2.20}) into Eq. (\ref{ap1.1}) directly leads to Eq.\ (\ref{2.14}).
 
To derive the balance equation for the energy, a similar procedure is followed. The Boltzmann equation is multiplied by $mv^{2}/2$ and afterwards integrated over the velocity. Using the balance equation for the momentum,  Eq. (\ref{2.14}), it is found
\begin{eqnarray}
\label{ap1.12}
\frac{3}{2} n({\bm r},t) \frac{\partial T({\bm r},t)}{\partial t} & + & \frac{3}{2} n({\bm r},t)) {\bm u}({\bm r},t) \cdot \frac{\partial T({\bm r},t)}{\partial {\bm r}} 
+{\sf P} ^{(k)} ({\bm r},t) : \frac{\partial {\bm u}({\bm r},t)}{\partial {\bm r}} \nonumber \\
&+& \frac{\partial}{\partial {\bm r}}\, \cdot {\bm J}_{q}^{(k)}({\bm r},t)- u_{i}({\bm r},t) \frac{\partial}{\partial z} P_{zi}({\bm r},t)  = \int d{\bm v} \frac{mv^{2}}{2} J[{\bm r},{\bm v}|f],
\end{eqnarray}
with the  the kinetic part of the heat flux, ${\bm J}_{q}^{(k)}$, given by Eq. (\ref{2.19}). Here, the rule of implicit sum over repeated indexes is used. To compute the integral involving the collision term, the general property given in Eq.\ (\ref{2.7}) is employed. From the collision rules it follows that
\begin{equation}
\label{ap1.13}
v^{\prime 2}-v^{2} = - \frac{1-\alpha^{2}}{4} ({\bm g} \cdot \widehat{\bm \sigma})^{2}+(1+\alpha) {\bm g} \cdot \widehat{\bm \sigma} {\bm G} \cdot \widehat{\bm \sigma}\, ,
\end{equation}
where ${\bm G}$ is the center of mass velocity,
\begin{equation}
\label{ap1.14}
{\bm G} \equiv \frac{{\bm v}+{\bm v}_{1}}{2}.
\end{equation}
In this way it is easily obtained that
\begin{equation}
\label{ap1.15}
\int d {\bm v} \frac{mv^{2}}{2} J[{\bm r},{\bm v}|f] =- (1-\alpha^{2} ) \omega ({\bm r},t)+I({\bm r},t).
\end{equation}
The expression of the source term $\omega ({\bm r},t)$ is given in Eq.\, (\ref{2.29}), while
\begin{eqnarray}
\label{ap1.16}
I({\bm r},t)  & = & -\frac{m\sigma}{2}\, (1+\alpha) \int d{\bm v} \int d {\bm v}_{1} \int_{\sigma/2}^{h-\sigma/2} dz_{1} \int_{0}^{2\pi} d \varphi\,  |{\bm g} \cdot \widehat{\bm \sigma}|^{2} {\bm G} \cdot \widehat{\bm \sigma} \Theta \left( -{\bm g} \cdot \widehat{\bm \sigma} \right) \nonumber \\
&& \times  f({\bm r}_{1z},{\bm v}_{1} ,t) f({\bm r},{\bm v},t).
\end{eqnarray}
By using the same kind of method as for the pressure tensor this can be written as
\begin{eqnarray}
\label{ap1.17}
I({\bm r},t)  & = & - \frac{m \sigma^{2}}{4}\ (1+\alpha) \frac{\partial}{\partial z}  \int d{\bm v} \int d {\bm v}_{1} \int_{\sigma/2}^{h-\sigma/2} dz_{1} \int_{0}^{2\pi} d \varphi \int_{0}^{1} d\lambda\, \lambda^{-1}   |{\bm g} \cdot \widehat{\bm \sigma}_{\lambda}|^{2} {\bm G} \cdot \widehat{\bm \sigma}_{\lambda}  \widehat{\sigma}_{\lambda z} \nonumber \\
&& \times  \Theta \left( -{\bm g} \cdot \widehat{\bm \sigma}_{\lambda} \right)  f({\bm r}_{1z},{\bm v}_{1} ,t) f({\bm r}_{\lambda},{\bm v},t).
\end{eqnarray}
Now, the peculiar velocity of the center of mass $ \mathcal{\bm G}$ defined in Eq.\ (\ref{2.24}) is introduced, and it is a simple task to get
\begin{equation}
\label{ap1.18}
I({\bm r},t)= -\frac{\partial}{\partial z}\, \left[ u_{i}({\bm r},t) {\sf P}_{zi}^{(c)}({\bm r},t)\right] -\frac{\partial}{\partial z} J_{qz}^{(c)}({\bm r},t).
\end{equation}
The expression of the collisional transfer contribution to the heat flux, $J_{qz}^{(c)}$ is given in Eq.\ (\ref{2.23}). To get Eq. (\ref{2.15}) in the main text,  it only remains to substitute the above result into Eq.\, (\ref{ap1.15}), and then use the derived expression for the collision term contribution into Eq.\ (\ref{ap1.12}).

 \section{The ICS distribution from the kinetic model equation}
 \label{ap2}
 
 In this Appendix, the derivation of the results reported at the beginning of Sec.\ \ref{s4} will be outlined. For the Maxwellian distribution in Eq.\ (\ref{4.1}), it is
 \begin{equation}
 \label{ap2.1}
 {\sf P}^{(k)} (z,t) =n(z) T(t) {\sf I},
 \end{equation}
 whwre ${\sf I}$ is the unit tensor in three dimensions. The kinetic part of the heat flux, ${\bm J}_{q}^{(k)}$, defined in Eq.\ (\ref{2.19}), vanishes since the distribution function  of the ICS  is an even function of the velocity.  
 
 From Eqs.\ (\ref{2.20}) and (\ref{ap1.2}),
 \begin{eqnarray}
 \label{ap2.2}
 \frac{\partial}{\partial z}\, {\sf P}_{zi}^{(c)}(z,t) & =  &\frac{1+\alpha}{2}\, m \sigma n(z) \int  d{\bm v} \int d {\bm v}_{1} \int_{\sigma/2}^{h-\sigma/2} dz_{1} \int_{0}^{2\pi} d \varphi\,  |{\bm g} \cdot \widehat{\bm \sigma} |^{2} \widehat{\sigma} _{i} \nonumber \\
 && \times \Theta  \left( - {\bm g} \cdot \widehat{\bm \sigma} \right) n(z_{1})   \psi_{ICS}({\bm v},t) \psi_{ICS}({\bm v}_{1},t) \nonumber \\
 &=& \frac{1+\alpha}{4}\, m \sigma n(z) \int  d{\bm v} \int d {\bm v}_{1} \int_{\sigma/2}^{h-\sigma/2} dz_{1} \int_{0}^{2\pi} d \varphi\,  |{\bm g} \cdot \widehat{\bm \sigma} |^{2} \widehat{\sigma}_{i}  \nonumber \\
 && \times  n(z_{1})   \psi_{ICS}({\bm v},t) \psi_{ICS}({\bm v}_{1},t).
 \end{eqnarray}
The velocity integrals are easily carried out to get
\begin{eqnarray}
\label{ap2.3}
\frac{\partial}{\partial z}\, {\sf P}_{zi}^{(c)}(z,t) &= &\frac{1+\alpha}{2} \sigma n(z) T(t) \int_{\sigma/2}^{h-\sigma/2}dz_{1} \int_{0}^{2\pi} d\varphi\, \widehat{ \sigma}_{i} n(z_{1}) \nonumber \\
& = &\delta_{iz}  \pi (1+\alpha) n(z) T(t) \int_{\sigma/2}^{h-\sigma/2} dz_{1}\, (z_{1}-z) n(z_{1}).
\end{eqnarray}

Finally, for the collisional transfer contribution to the heat flux, use of the distribution (\ref{4.1}) into Eq.  (\ref{2.23}) gives
\begin{eqnarray}
\label{ap2.3}
J_{q,z}^{(c)}(z,t) &= &\frac{m \sigma^{2}}{8} (1+\alpha) n(z) \int d{\bm v} \int d {\bm v}_{1}  \int_{\sigma/2}^{h-\sigma/2} dz_{1} \int_{0}^{2\pi} d \varphi \int_{0}^{1} d \lambda\,  |{\bm g} \cdot \widehat{\bm \sigma}_{\lambda}|^{2}  \nonumber \\
& & \times {\bm G} \cdot \widehat{\bm \sigma}_{\lambda} \widehat{\sigma}_{\lambda z} \psi_{ICS}({\bm v},t) \psi_{ICS}({\bm v}_{1},t),
\end{eqnarray}
and this expression vanishes because the integrand is and even function of the velocities. Taking into accout the above results, particularization of Eq.\ (\ref{3.16}) for $f_{ICS}(z,{\bm v},t)$, given in Eq. (\ref{4.1}), yields
\begin{equation}
\label{ap2.4}
\frac{\partial f_{ICS}}{\partial t}+ v_{z} \frac{\partial f_{ICS}}{\partial z} = -\frac{ \psi_{ICS}}{T(t)} \left[ v_{z} \frac{\partial {\sf P}_{z}^{(c)}}{\partial z} + \left( \frac{mv^{2}}{3T(t)} -1 \right) (1-\alpha^{2} ) \omega \right]
\end{equation}
or, equivalently,
\begin{equation}
\label{ap2.5}
\frac{3}{2T(t)}\, n(z) \frac{\partial T(t)}{\partial t} \left( \frac{mv^{2}}{3T(t)} -1 \right) - v_{z} \frac{\partial n(z)}{\partial z} = \frac{1}{T(t)} \left[ v_{z} \frac{\partial {\sf P}_{z}^{(c)}}{\partial z} + \left( \frac{mv^{2}}{3T(t)} -1 \right) (1-\alpha^{2} ) \omega \right].
\end{equation}
This equation directly implies Eqs.\ (\ref{4.3}) and (\ref{4.4}), just by equating coefficients of the same power of the velocity.

\end{document}